\documentclass[eqsecnum,aps,showpacs,amsmath,nofootinbib,twocolumn, superscriptaddress,preprintnumbers]{revtex4}
\usepackage{dcolumn}

\usepackage{epsfig, color,graphicx}
\usepackage{amsmath}
\usepackage{amssymb}
\begin{document}


\title{Wormholes in higher dimensional space-time: \\
Exact solutions and their linear stability analysis
}
\author{Takashi Torii}\email{torii@ge.oit.ac.jp}
\affiliation{Department of General Education, Osaka Institute of Technology,
Asahi-ku, Osaka 535-8585, Japan}

\author{Hisa-aki Shinkai}\email{shinkai@is.oit.ac.jp}
\affiliation{Department of Information Systems, Osaka Institute of Technology,
Kitayama, Hirakata, Osaka 573-0196, Japan}
\affiliation{Computational Astrophysics Laboratory,
 Institute of Physical \& Chemical Research (RIKEN),\\
 Hirosawa, Wako, Saitama, 351-0198 Japan
}

\date{\today 
}

\begin{abstract}
We derive the simplest traversable wormhole solutions in $n$-dimensional general relativity, 
assuming static and spherically symmetric spacetime with 
a ghost scalar field. 
This is the generalization of 
the Ellis solution (or the so-called Morris-Thorne's traversable wormhole)
into a higher-dimension. 
We also study their stability using linear 
perturbation analysis. 
We obtain the master equation for the perturbed gauge-invariant variable 
and search their eigenvalues. 
Our analysis shows that all higher-dimensional wormholes have an unstable mode
against the perturbations with which the throat radius is changed.
The instability is consistent with the earlier numerical analysis in 
four-dimensional solution.  




\end{abstract}



\maketitle

\section{Introduction}
\label{section_intro}
Wormholes are popular tools in science fiction as a way 
for rapid interstellar travel, time machines and warp drives.  
However, wormholes are also a scientific topic,  
just after the birth of general relativity. 

Historically, a ``tunnel structure" in the Schwarzschild solution
was first pointed out by Flamm in 1916\cite{Flamm1916}.  
Einstein and Rosen \cite{EinsteinRosen} proposed a ``bridge structure"
between black holes in order to obtain a regular solution without a singularity. 
The name ``wormhole" was coined by John A. Wheeler in 1957,  
and its fantastic applications are popularized 
after the influential study of traversable wormholes by
Morris and Thorne \cite{MT}.  

Morris and Thorne  considered ``traversable conditions" for human travel 
through wormholes responding to Carl Sagan's idea for his novel ({\it Contact}),  
and concluded that such a wormhole solution is available 
if we allow ``exotic matter" (negative-energy matter). 

The introduction of exotic matter sounds to be unusual for the first time, but 
such matter appears in quantum field theory and in alternative
gravitational theories such as scalar-tensor theories. 
The Morris-Thorne solution is constructed with a
massless Klein-Gordon field whose gravitational coupling takes the opposite
sign to normal, which appears in Ellis's earlier work \cite{Ellis}, 
who called it a drainhole, 
and also in more general framework of scalar-tensor
theories by Bronnikov in the same year\cite{Bronnikov1973}. 
(See a review e.g. by Visser \cite{MVbook} for earlier works; See also e.g. Lobo
\cite{Lobobook} for recent works).

Since the difference of light bending behavior between the Ellis wormhole and 
Schwarzschild black hole were reported by Abe\cite{abe2010}, the microlensing images
with wormholes are also getting attention from the observational point of view \cite{toki,yoo}.


One of our main motivations in this paper is the dynamical features of wormholes.
A wormhole is supposed to connect two spacetimes as a two-way interface, while
a black hole is an one-way interface. 
From this analogy, Hayward \cite{Hayward1999} proposed a unified understanding 
of black holes and traversable wormholes, {\it i.e.} a wormhole 
throat can be interpreted as a degenerate horizon.
This idea predicts that a wormhole changes to a black hole in its dynamical 
evolutions in the classical process. 

This is numerically shown by one of the authors \cite{ShinkaiHayward}.
Using a dual-null formulation for space-time integration, 
they observed that the wormhole is unstable against
Gaussian pulses in either an exotic or normal massless Klein-Gordon field. 
The wormhole throat suffers a bifurcation of the horizon and either explodes to form
an inflationary universe or collapses to a black hole, whether the total input
energy is negative or positive, respectively. 

These basic behaviors were repeatedly confirmed by other groups
\cite{Doroshkevich,Sarbach2}, together with linear perturbation analysis \cite{Sarbach1} 
\footnote{
Armendariz-Picon\cite{Armendariz-Picon} reported that the Ellis wormhole is stable
using perturbation analysis. 
However, Gonzalez et al \cite{Sarbach1} reported that 
his conclusion is within the limited class of perturbations and
the Ellis wormhole is unstable}. 
The wormhole solutions with a conformal scalar field were reported\cite{Bronnikov1973,BarceloVisser}, and their instabilities are shown also 
using linear perturbation analysis
\cite{BronnikovGrinyok2004}. 
There are also discussions on the wormhole solutions in alternative/modified 
gravity (e.g. \cite{BKZ2012, NovikovShatskiy}). 
Wormhole thermodynamics is also proposed based on these properties \cite{Hayward2009}.

We, therefore, understand that four-dimensional Ellis wormhole is unstable. 
If this feature can also be seen in higher-dimensional spacetime, 
it should be generic independent of the dimension.
The higher-dimensional theories such as string/M theories are applied for various unsolved problems in gravitational phenomena and cosmology, and we gain new insights into them.
The wormholes in higher dimensional general relativity 
lead to the study in such fundamental theories.

Wormhole study in higher-dimensional spacetime is not a new topic. 
We can find the articles from 1980s 
\cite{ChadosDetweiler,Clement84}, and 
the recent studies include higher-curvature terms (see e.g. \cite{MaedaNozawa}
and \cite{KKK} and references therein).
Most of the research concerns the solutions and their energy conditions mainly, 
but to our knowledge there is no general discussion on the stability analysis of the
solutions. 
 




In this article, we 
construct Ellis solutions in higher-dimensional general relativity, 
and study their stability using the linear perturbation technique. 
The full numerical studies will be shown in our follow up paper. 

This paper is organized as follows. 
In Section~\ref{section_2}, we derive our higher-dimensional wormhole solutions. 
In Section~\ref{section_3}, we show the linear perturbation analysis. 
The conclusion and discussion are shown in Section~\ref{section_4}.

\section{Wormhole solutions}
\label{section_2}
We start from the $n$-dimensional Einstein-Klein-Gordon system
\begin{eqnarray}
S = \int \! d^{n}x \sqrt{-g} \bigg[\frac{1}{2\kappa_n^2}R 
-\frac{1}{2} \epsilon (\nabla \phi)^2-V(\phi) \bigg],
\label{action1}
\end{eqnarray}
where $\kappa_n^2$ is a $n$-dimensional gravitational constant.
The scalar field $\phi$ can be called as {\it normal} (or {\it ghost}) field 
if $\epsilon=1\; (-1)$. 

This action derives the Einstein equation 
\begin{eqnarray}
G_{\mu\nu}=\kappa_n^2 T_{\mu\nu}, 
\end{eqnarray}
where 
\begin{eqnarray}
T_{\mu\nu} = \epsilon(\partial_{\mu}\phi)(\partial_{\nu}\phi)-g_{\mu\nu}\Bigl[\frac{1}{\:2\:}\epsilon (\nabla\phi)^2+V(\phi)\Bigr], \label{Tmunu}
\end{eqnarray}
and the Klein-Gordon equation
\begin{eqnarray}
\kern1pt\vbox{\hrule height 0.9pt\hbox{\vrule width
0.9pt\hskip 2pt\vbox{\vskip 5.5pt}\hskip 3pt\vrule width 0.3pt}\hrule height
0.3pt}\kern1pt
\phi  =-\epsilon \frac{dV}{d\phi}.
\label{KG}
\end{eqnarray}

We consider the 
space-time with the metric 
\begin{eqnarray}
ds^2 &=& - f(t,r)e^{-2\delta(t,r)}dt^2 + f(t,r)^{-1} dr^2 \nonumber \\ 
&& + R(t,r)^2 h_{ij}dx^i dx^j,
\label{metric}
\end{eqnarray}
where $h_{ij}dx^i dx^j$ represents the line element of a unit $(n-2)$-dimensional
constant curvature space with curvature $k=\pm1,\:0$ and volume $\Sigma_k$.


In order to construct a static wormhole solution, we restrict the metric function 
as $f=f(r)$, $R=R(r)$, $\phi=\phi(r)$, and $\delta=0$.
The $(t,t)$,  $(r,r)$, and  $(t,r)$ components of the Einstein equations, then, become
\begin{widetext}
\begin{eqnarray}
&&-\frac{n-2}{2}f^2\biggl[\frac{2R''}{R}+\frac{f'R'}{fR}+\frac{(n-3)R'^2}{R^2}\biggr] +\frac{(n-2)(n-3)kf}{2R^2}
=\kappa_n^2f\Bigl[ \frac{1}{\:2\:}\epsilon f \phi'^2+V(\phi)\Bigr], 
\rule[0mm]{0mm}{8mm} 
\label{Einstein_static_tt}\\
&&\frac{n-2}{2}\frac{R'}{R}\biggl[\frac{f'}{f}+\frac{(n-3)R'}{R}\biggr] -\frac{(n-2)(n-3)k}{2fR^2}
=\frac{\kappa_n^2}{f}\Bigl[\frac{1}{\:2\:}\epsilon f \phi'^2-V(\phi)\Bigr],
\rule[0mm]{0mm}{8mm} \\
&&\frac{f''}{2}+(n-3)f\biggl(\frac{R''}{R}+\frac{f'R'}{fR}+\frac{n-4}{2}\frac{{R'}^2}{R^2}\biggr)
-\frac{(n-3)(n-4)k}{2R^2}
=\kappa_n^2\Bigr[\frac{1}{\:2\:}\epsilon f \phi'^2+V(\phi)\Bigr],
\rule[0mm]{0mm}{8mm}
\label{Einstein_static_ij}
\end{eqnarray}
\end{widetext}
respectively, and the Klein-Gordon equation becomes
\begin{eqnarray}
\frac{1}{R^{n-2}}\big(R^{n-2}f\phi'\big)' 
=-\epsilon \frac{dV}{d\phi}. 
\label{KG-2}
\end{eqnarray}

Hereafter, we assume that the scalar field is ghost ($\epsilon=-1$)
and massless ($V(\phi)=0$).
The Klein-Gordon equation (\ref{KG-2}) is integrated as 
\begin{eqnarray}
\phi' = \frac{C}{fR^{n-2}}, 
\label{masless-KG-1}
\end{eqnarray}
where $C$ is an integration constant.
The Einstein equations 
(\ref{Einstein_static_tt})--(\ref{Einstein_static_ij}) are reduced to 
\begin{eqnarray}
\frac{(n-2)R'}{R}\Big[\frac{f'}{f}+\frac{(n-3)R'}{R}\Big]
-\frac{(n-2)(n-3)k}{fR^2} \nonumber \\
=-\frac{\kappa_n^2 C^2}{f^2R^{2(n-2)}}
\label{masless-eq-1}
\end{eqnarray}
and
\begin{eqnarray}
\frac{(n-2)R''}{R}=\frac{\kappa_n^2 C^2}{f^2R^{2(n-2)}}.
\label{masless-eq-2}
\end{eqnarray}

We assume the throat of the wormhole is at $r=0$, and 
$a$ is the radius of the throat, i.e., $R(0)=a$. 
By the regularity conditions at the throat,   
\begin{eqnarray}
R(0)=a>0, \;\mbox{and}\; f(0)=f_0>0,
\label{regularity}
\end{eqnarray}
where $f_0$ is a constant. Here we can assume $a=1$ and $f_0=1$ without loss of generality \cite{note_1},
but we keep $a$ in the equations in this section for later convenience.
We also assume the reflection symmetry with respect to the throat: 
\begin{eqnarray}
R'(0)=0, \;\mbox{and}\; f'(0)=0.
\label{reflection}
\end{eqnarray}
There is a shift symmetry of the scalar field $\phi$ and we impose $\phi(0)=0$. 
By substituting these conditions into Eq.~(\ref{masless-eq-1}), 
the integration constant $C$ is determined as 
\begin{eqnarray}
\kappa_n^2 C^2 = (n-2)(n-3)ka^{2(n-3)}.
\label{constant_C}
\end{eqnarray}
For the case $k=0$, the constant $C$ vanishes and the solution becomes
trivial. For the case $k=-1$, Eq.~(\ref{constant_C}) is not satisfied and there is no wormhole solution. Below we assume  $k=1$.

The solution of Eqs.~(\ref{masless-KG-1})--(\ref{masless-eq-2}) is obtained as 
\begin{eqnarray}
f&\equiv& 1, 
\label{massless-sol-1}
\\
R' &=& \sqrt{1-\Big(\frac{a}{R}\Big)^{2(n-3)}}, 
\label{massless-sol-2}
\\
\phi&=& \frac{\sqrt{(n-2)(n-3)}}{\kappa_n} a^{n-3}  \int\frac{1}{R(r)^{n-2}}dr.
\label{massless-sol-3}
\end{eqnarray}
The Eq.~(\ref{massless-sol-2}) is integrated to give
\begin{eqnarray}
r(R)=-mB_{z}\Big(\!-m,\: \frac{1}{\:2\:} \Big)
-\frac{\sqrt{\pi}\Gamma[1-m]}{\Gamma[m(n-4)]},
\label{massless-sol-4}
\end{eqnarray}
where $m={1}/{2(n-3)}$ and $z=R^m$. $B_{z}(p,q)$ is the incomplete beta function defined by
\begin{eqnarray}
B_{z}(p, q):= \int_0^{z} t^{p-1}(1-t)^{q-1} dt
\end{eqnarray}
which can be expressed by the hypergeometric function $F(\alpha, \beta, \gamma; z)$ as
\begin{eqnarray}
B_{z}(p, q)=\frac{z^p}{p}F(p,\: 1-q,\: p+1;\: z).
\end{eqnarray}
Although Eq.~(\ref{massless-sol-4}) is implicit with respect to $R$, it is rewritten in the explicit form
by using the inverse incomplete beta function.
For $n=4$, this solution reduces to Ellis's wormhole solution.
\begin{equation}
f\equiv 1, ~~~
R=\sqrt{r^2+a^2}, ~~~
\phi=\sqrt{2}\tan^{-1} \frac{r}{a}.
\end{equation}

At the throat, we find
\begin{eqnarray}
R''(a)=\frac{n-3}{a}, \mbox{~and~}
\phi'(a)=\frac{\sqrt{(n-2)(n-3)}}{\kappa_n a}.
\end{eqnarray}
These indicate that the throat of the wormhole has larger curvature
and the scalar field $\phi$ becomes steeper as $n$ goes higher. 
At the spacial infinity, 
the scalar field $\phi(r)$ becomes constant and the function $R(r)$ is proportional to $r$.
We plotted these behaviors in Figure~\ref{fig1}. 
For $n\to \infty$, the functions have the limiting solution, 
$R= r+a$ and $\phi = \pi/2~(r>0)$.

\begin{widetext}
\begin{center}
\begin{figure}[tb]
\includegraphics[width=16cm]{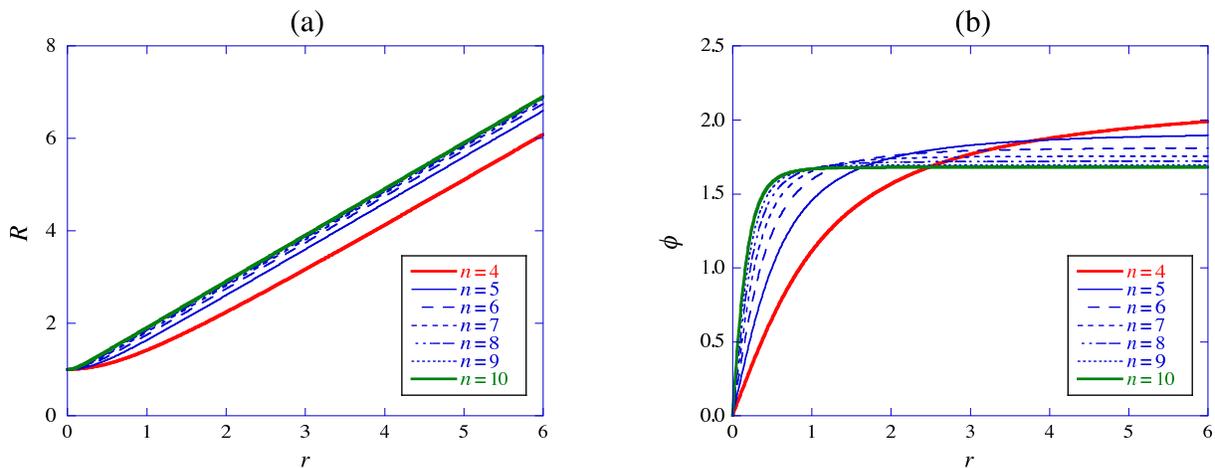}
\caption{The $n$-dimensional wormhole solutions;  
(a) The circumference radius $R$ and (b) the scalar field $\phi$ are plotted as a function of the radial coordinate $r$. The cases of $n=4$--$10$ are shown.}
\label{fig1}
\end{figure}
\end{center}
\end{widetext}

\section{Stability analysis}
\label{section_3}
In this section, we investigate the linear stability of the higher-dimensional wormhole
solution obtained in the previous section.
In the non-linear analysis in four-dimensional spacetime, it is shown that the 
instability occurs
by resolution of the degeneracy of a double trapping horizon by perturbing the 
throat radius
\cite{ShinkaiHayward}. 
Hence we follow the analysis in Ref.~\cite{Sarbach1},
where the throat radius is not fixed.

We focus on the ``spherical" modes, where the $(n-2)$-dimensional constant curvature space
is not perturbed \cite{spherical}. 
In the time-dependent and spherically symmetric spacetime, the metric is written as 
Eq.~(\ref{metric}) generally.  We write the perturbed functions as 
\begin{eqnarray}
f(t,r)\:=\!\!&&\!\!f_0(r)+\varepsilon f_1(r)e^{i\omega t}, \\
\delta(t,r)\:=\!\!&&\!\!\delta_0(r)+\varepsilon\delta_1(r)e^{i\omega t}, \\
R(t,r)\:=\!\!&&\!\!R_0(r)+\varepsilon R_1(r)e^{i\omega t}, \\
\phi(t,r)\:=\!\!&&\!\!\phi_0(r)+ \varepsilon\phi_1(r)e^{i\omega t}.
\end{eqnarray}
$\varepsilon$ is an infinitesimal parameter. The variables with subscript 0 denote the static solution obtained in the previous section.
This ansatz contains one gauge mode.

\begin{widetext}
The first-order equations of the Einstein equations become
\begin{eqnarray}
&&
R_1''
+\frac{(n-3) R_0'}{R_0}R_1'
+\frac{R_0'}{2}f_1'
+\frac{(n-3)}{2 R_0}f_1
-\sqrt{\frac{n-3}{n-2}} \frac{1}{R_0^{n-3}}\phi_1'
=0,
\label{lin_tt}
\\ 
&&
\frac{(n-3) R_0'}{R_0}R_1'
+\frac{n-3}{R_0^{2 n-4}}R_1
+\frac{R_0'}{2}f_1'
+\frac{(n-3)}{2 R_0}f_1
-R_0' \delta_1'
+\sqrt{\frac{n-3}{n-2}} \frac{1}{R_0^{n-3}}\phi_1'
+R_1 \omega ^2
=0,
\label{lin_rr}
\\ 
&&
2R_1'
-R_0'f_1
-2 \sqrt{\frac{n-3}{n-2}} \frac{1}{R_0^{n-3}}\phi_1
=0,
\label{lin_rt}
\end{eqnarray}
for the $(t,t)$, $(r,r)$, and $(t,r)$ components, respectively. 
Here we assume $a=1$.
{}From  Eq.~(\ref{lin_rt}),  $f_1$ is
\begin{eqnarray}
f_1=
2\sqrt{\frac{n-3}{n-2}}\frac{1}{R_0^{n-3}R_0'}\phi_1 
-\frac{2}{R_0'}R_1'.
\label{eq_f1}
\end{eqnarray}
%
By substituting Eq.~(\ref{eq_f1}) into Eqs.~(\ref{lin_tt}) and (\ref{lin_rr}),  we find
\begin{eqnarray}
R_1''
-\frac{n-3}{R_0^{2 n-4}}R_1
+R_0' \delta_1'
-2\sqrt{\frac{n-3}{n-2}}\frac{1}{R_0^{n-3}}\phi_1' 
=\omega ^2R_1.
\label{eq_R1rr}
\end{eqnarray}
With Eq.~(\ref{eq_f1}), the Klein-Gordon equation turns out to be
\begin{eqnarray}
&&\phi_1''
+\frac{(n-2)+(n-4) R_0^{-2n+6}}{R_0R_0'}\phi_1'
-\frac{2 (n-3)^2}{R_0^{2 n-4}R_0'^2}\phi_1
-\frac{2 \sqrt{(n-2)_3}}{R_0^{n-2}R_0'}R_1'' 
\nonumber
\\
&&~~~~~+\frac{\sqrt{(n-2)_3}\bigl[(n-2)+(n-4) R_0^{-2 n+6}\bigr]}{R_0^{n-1}R_0'^2} R_1'
-\frac{(n-2) \sqrt{(n-2)_3}R_0' }{R_0^{n}} R_1
-\frac{\sqrt{(n-2)_3}}{R_0^{n-2}}\delta_1' 
=
-\omega^2\phi_1.
\label{eq_phi1rr}
\end{eqnarray}

\end{widetext}

By introducing the new variable, 
\begin{eqnarray}
\psi_1=R_0^{\frac{n-2}{2}}\Bigl(\phi_1-\frac{\phi_0'}{R_0'}R_1\Bigr),
\end{eqnarray}
we find Eqs.~(\ref{eq_R1rr}) and (\ref{eq_phi1rr}) give the single master equation, 
\begin{eqnarray}
-\psi_1''+V(r)\psi_1=\omega^2 \psi_1, 
\label{master_1}
\end{eqnarray}
with the potential, 
\begin{equation}
V(r)=
\frac{n-2}{2}\biggl[\frac{n-3}{R_0^{2(n-2)}}+\frac{(n-4)R_0'^2}{2R_0^2}\biggl]
+\frac{2(n-3)^2}{R_0^{2(n-2)}{R_0'}^2}. \label{potentialfunc}
\end{equation}
The variable 
$\psi_1$ is gauge invariant under the spherically symmetric ansatz.  However, 
$R_0'$ is zero at the throat, and the potential $V$ diverges there. 
Hence we regularize the master equation (\ref{master_1})\cite{note_2}.

\begin{figure}[t]
\includegraphics[width=7.5cm]{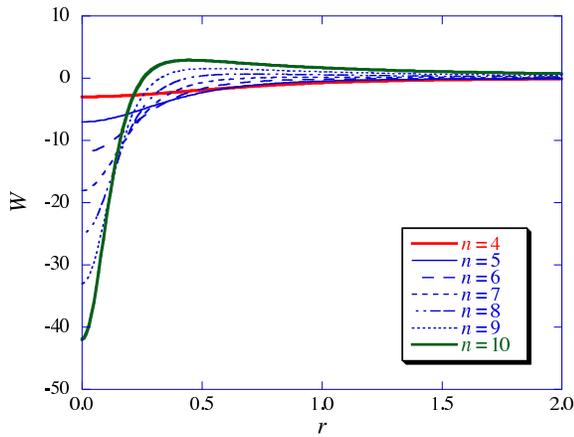}
\caption{
The potential function $W(r)$ is plotted. 
$W(r)$ is finite everywhere, and negative around the throat.
}
\label{fig_pot}
\end{figure}

It is easily checked that the master equation (\ref{master_1}) has a 0-mode solution 
\begin{eqnarray}
\bar{\psi}_1=\frac{1}{R_0^{\frac{n-4}{2}}R_0'}.
\label{0-mode}
\end{eqnarray}
With the 0-mode solution, we define differential operators
\begin{eqnarray}
{\cal D}_{+}=\dfrac{d}{dr}-\frac{\bar{\psi}_1'}{\bar{\psi}_1}
\quad \mbox{and} \quad 
{\cal D}_{-}=-\dfrac{d}{dr}-\frac{\bar{\psi}_1'}{\bar{\psi}_1}.
\label{dif-ope}
\end{eqnarray}
Then the master equation, (\ref{master_1}), can be written as
\begin{eqnarray}
{\cal D}_{-}{\cal D}_{+}\psi_1=\omega^2 \phi_1.
\label{master_2}
\end{eqnarray}
Operating ${\cal D}_{+}$ from the left and defining the new variable $\Psi_1={\cal D}_{+}\psi_1$,
we find the regularized master equation
\begin{eqnarray}
-\Psi_1''+W(r)\Psi_1=\omega^2 \Psi_1,
\label{master_3}
\end{eqnarray}
where 
\begin{equation}
W(r)=-\frac{1}{4R_0^2}\Bigl[\frac{3(n-2)^2}{R_0^{2(n-3)}}-(n-4)(n-6)\Bigr].
\label{regularpot}
\end{equation}
Figure~\ref{fig_pot} shows the configurations of the potential function $W(r)$.
Now the potential function is regular everywhere.
For $n=4$, $W(r)$ has the minimum at the throat and is negative definite.
For $n\geq 5$, $W(r)$ has the minimum at the throat, while it increases apart from the throat and becomes 
positive for large $r$.

\begin{figure}[th]
\includegraphics[width=7.5cm]{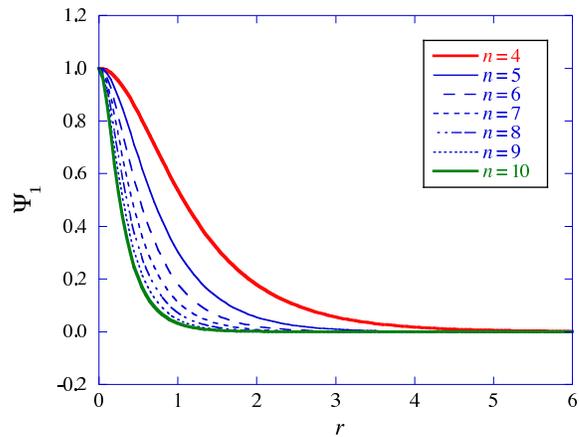}
\caption{
The eigenfunction $\Psi_1$ [Eq.~(\ref{master_3})] is plotted as a function of radial coordinate $r$. 
}
\label{fig_psi}
\end{figure}

We search the eigenfunctions $\Psi_1(r)$ of Eq.~(\ref{master_3}), and
find them in any dimension $n$. There exists one negative eigenvalue 
for $\omega^2$, which are listed in Table~\ref{table-eigenvals}. 
The existence of the eigenfunction with negative $\omega^2$ implies that
the solution is unstable. 
We find large negative $\omega^2$ for higher $n$, which indicates the time-scale of
instability becomes shorter.  This feature corresponds to the depth of the potential $W$. 
The associated eigenfunctions $\Psi_1(r)$ are shown in Figure~\ref{fig_psi}.

\begin{table}[t]
\caption{The negative eigenvalues $\omega^2$.
\label{table-eigenvals}}
\begin{tabular}{rr}
\hline
$n$ & $\omega^2$~~~~~~~~~~ \\
\hline
4&$-1.39705243371511$\\
5&$-2.98495893027790$\\
6&$-4.68662054299460$\\
7&$-6.46258414126318$\\
8&$-8.28975936306259$\\
9&$-10.1535530451867~\:$\\
10&$-12.0442650147438~\:$\\
11&$-13.9552091676647~\:$\\
20&$-31.5751101285105~\:$\\
50&$-91.3457759137153~\:$\\
~~100&$~~-191.283017729717~~~\:\!$\\
\hline
\end{tabular}
\end{table}



\section{Conclusions and Discussions
}
\label{section_4}
%
We derived the simplest wormhole solutions 
in higher-dimensional general relativity.
The spacetime is assumed to be static and spherically symmetric, 
has ghost scalar field, 
and has reflection symmetry at the throat.  
The four-dimensional version is known as the Ellis (Morris-Thorne) 
solution. 
At the throat, both the ingoing and outgoing expansions vanish, which means 
that the throat consists of a degenerate horizon. 

The obtained solutions are expressed with the incomplete beta function. 
We expect that the solution can be expressed more simple functional form if we use
another coordinate system.  Or such an expression might have appeared in the 
literature, but we have not noticed it.  
However, we believe the successive stability analysis is new to us.

From the stability analysis using the linear perturbation technique, 
we showed that the solutions have one negative mode, 
which concludes that 
all wormholes are linearly unstable.
The time scale of instability becomes shorter as $n$ becomes large. 

By extrapolating the knowledge of four-dimensional Ellis's wormhole, we expect 
that these higher-dimensional wormholes also change to a black hole or 
an expanding throat.  This is actually true.  
In our succeeding papers, we will 
report the numerical evolutions of higher-dimensional wormholes,  
in which we show the above predictions are realized. 
Both linearly perturbed solutions and solutions with nonlinear pulse input 
suffer the bifurcations of horizons and turn to either black hole or expanding throat. 
In order to obtain a robust wormhole solution for such a disturbance, we may have to work 
in modified gravity theories, as was recently reported in dilaton-Gauss-Bonnet gravity\cite{KKK}. 

The instability of wormholes requires additional maintenance techniques in science fiction. 
Not only so, but this indicates that such a simple wormhole construction cannot be available as an 
astrophysical object with the present setting.

\section*{Acknowledgements}
T.T. thanks Ken Kamano, Hideki Maeda, Umpei Miyamoto, Masaaki Morita and Makoto Narita for useful discussions. 
We also thank the anonymous referee for pointing out earlier references. 
This work was supported in part by the Grant-in-Aid for
Scientific Research Fund of the JSPS (C) No. 22540293, and (C) No. 25400277. 
Numerical computations were carried out on SR16000 at YITP in Kyoto University, 
and on the RIKEN Integrated Cluster of Clusters (RICC).



\begin{thebibliography}{99}
\bibitem{note_1}
Introducing the new variables $\tilde{t}=t/a$, $\tilde{r}=r/a$, $\tilde{R}=R/a$, 
we can scale out the throat radius $a$.
\bibitem{spherical}
In the higher-dimensional spacetime, the constant curvature spaces with $k=1$ are not only 
the spherically symmetric but there are other spaces such as the one with Bohm metric.
However, here we will call the modes ``spherical".
\end{thebibliography}
\end{document}